\documentclass[a4paper]{jpconf}
\usepackage{graphicx}
\usepackage{url}
\begin{document}
\title{Temporal Variations of High-Degree Solar p-Modes using Ring-Diagram Analysis}

\author{Olga Burtseva$^1$, Sushant Tripathy$^1$, Richard Bogart$^2$, Kiran Jain$^1$, Rachel Howe$^3$, Frank
Hill$^1$,  and Maria Cristina Rabello-Soares$^4$}

\address{$^1$ National Solar Observatory, Tucson, AZ, USA}
\address{$^2$ HEPL, Stanford University, CA, USA}
\address{$^3$ School of Physics and Astronomy, Edgbaston, Birmingham, UK}
\address{$^4$ Departamento de F\'{\i}sica, Universidade Federal de Minas Gerais, Belo Horizonte, Brazil}

\ead{burtseva@noao.edu}

\begin{abstract}
We study temporal variations in the amplitudes and widths of high-degree acoustic modes by applying the ring-diagram technique to the GONG+, MDI and HMI Dopplergrams during the declining phase of cycle 23 and rising phase of cycle 24. The mode parameters from all three instruments respond similarly to the varying magnetic activity. The mode amplitudes and widths show consistently lower variation due to smaller magnetic activity in cycle 24 as compared to the previous solar cycle.   
\end{abstract}

\section{Introduction}
The structural and dynamical properties of the Sun, changing with the solar cycle, are reflected in variations of solar oscillation parameters. In this context, we study temporal variations of the amplitudes and widths of high-degree solar acoustic modes. Temporal variations in these parameters can give us insight into the processes of excitation and damping of the solar acoustic oscillations, and mechanisms regulating the solar magnetic cycle. The global and local analyses of solar acoustic modes have shown that the mode amplitudes and lifetimes are anti-correlated with the solar activity level, and strongly depend on the local magnetic flux \cite{Chaplin00,Komm00,Rajaguru01,Howe04}. In this work, we derive the amplitudes and widths of high-degree acoustic modes by applying the ring-diagram technique \cite{Hill88} to the GONG (Global Oscillation Network Group), MDI (Michelson Doppler Imager) and HMI  (Helioseismic and Magnetic Imager) Dopplergrams during the declining phase of cycle 23 and rising phase of cycle 24. We compare the results with level of magnetic activity in the analyzed regions.    

\section{Data analysis} 
The mode parameters of the solar acoustic oscillations in this work are obtained from GONG (for the period from 2001 to 2012), MDI (for the period from 2001 to 2010) and HMI (for the period from 2010 till 2013) Dopplergrams using the standard ring-diagram technique of the GONG and HMI pipelines. The size of a standard patch in the ring analysis is $15^\circ \times 15^\circ$. The GONG pipeline routine uses a symmetric Lorentzian profile \cite{Haber00} to fit the power spectrum of the solar oscillations. The HMI pipeline \cite{Bogart11} applies both symmetric and asymmetric \cite{Basu99} profiles to derive the mode
parameters. The results from both procedures are generally consistent for the parameters of interest here; thus, we show results from the symmetric profile fitting only.

The Magnetic Activity Index (MAI) values are computed from MDI 96-minute \cite{Basu04} and HMI 45-second magnetograms for the same location and times as the Doppler data. 

The amplitudes and widths fitted from the ring analysis need to be corrected for center-to-limb and duty-cycle dependencies \cite{Howe04}. In this work, we discuss results from disk center patches where the foreshortening effect is minimal. GONG data were restricted to time sequences with a duty cycle of 70$\%$ or higher and corrected for the duty-cycle dependence using a linear regression technique \cite{Komm00}. MDI data during Dynamics Runs usually have a high duty cycle. However the Dynamics Runs are restricted to about two months of observations per year. In this work, we also use MDI observations between the Dynamics Campaigns when available with a sufficiently high duty cycle in order to get a better temporal MDI coverage of the mode parameters. Most of these data have a lower duty cycle, so we also restrict the MDI data to times with a duty cycle of 70$\%$ or higher, and correct the mode parameters for the duty-cycle dependence. We analyzed 453 days from MDI, improving MDI data coverage in comparison with previous work \cite{Burtseva11}. HMI amplitudes and widths were corrected for the low duty-cycle related offsets during the HMI semi-annual daily eclipse periods.

\section{Results and discussion}  
The results for amplitudes and widths shown here are for a multiplet $\ell$ = 440, $n$ = 2 ($\nu$ = 3.2 mHz). The correlation of its variations over a solar activity cycle with other multiplets in the 2.5$-$3.5 mHz frequency range is above 70$\%$. 

The variations of amplitude and width as a function of time and MAI, computed from GONG, MDI and HMI data, are shown in Figure 1. The quietest period during the last solar minimum, according to the Mt.Wilson sunspot index data (\url{http://obs.astro.ucla.edu/150_data.html}) and MAI from MDI, was observed during most of the year 2008 and the beginning of 2009. However, HMI observations started in April of 2010; thus for consistency we plot the mode parameters from all three instruments relative to the mean value for the year 2010. 

\begin{figure}[t]
\begin{center}
\includegraphics[width=0.98\linewidth]{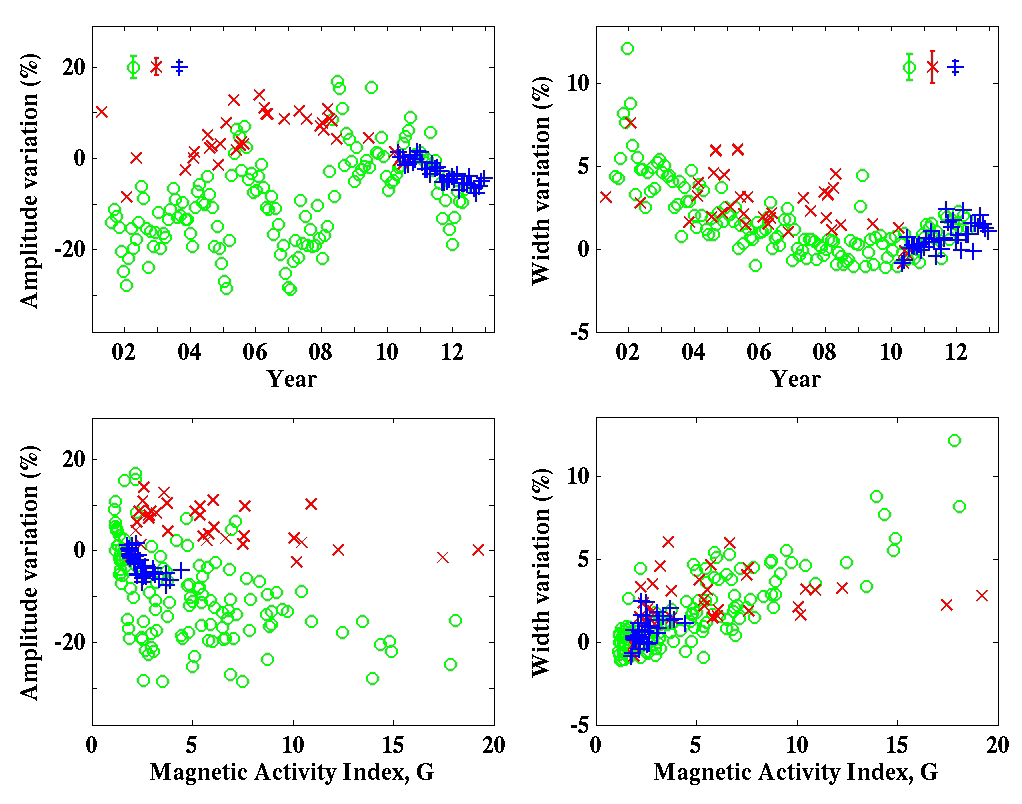}
\end{center}
\caption{Variation of amplitude and width (relative to the mean value of 2010) as a function of time and MAI, computed from GONG (open green cicrcles), MDI (red crosses) and HMI (blue pluses). The mode parameters are derived from disk center patches. Each point is an average over one Carrington rotation. The typical error bars for the measurements are shown in the top panels.}
\end{figure}

The amplitudes from GONG increase by $\sim 4\%$ from the high solar activity period in 2001 till 2004. From 2004 through 2008, we find a long-term variation of a non-magnetic nature (see, \cite{Burtseva11}). A co-temporal systematical problem was found in GONG velocity measurements, especially those from the Learmonth station. We believe this caused the long-term variations seen in the mode amplitudes. The mode widths are affected to a much smaller degree. They increase by an amount comparable with the error of the width measurement and are thus not obvious in the plot. We plan to explore the problem and correct it in a subsequent analysis. Ignoring these long-term variations, we conclude that the mode amplitudes obtained from GONG increased from 2001 to 2008-2009 by $\sim 18\%$, and then decreased from 2009 to 2012 by $\sim 10\%$. The widths decreased from 2001 to 2008 by $\sim 7\%$. A rising trend in the widths also starts from $\sim$ 2009 and reaches $\sim 2\%$ by mid-2012. After correction for duty cycle, the amplitude of the solar cycle variations for cycle 23 is somewhat lower than that reported in \cite{Burtseva11}. The overestimation of solar cycle variations was due to a lower temporal coverage of GONG high-resolution observations (and consequently lower amplitudes and higher widths) in the early years and then a gradual increase.

The variations in amplitude and width from MDI data show a solar cycle trend similar to that from GONG data. The mode amplitudes obtained from MDI from 2001 to 2008 increased by $\sim 15\%$ and the mode widths decreased by $\sim 4\%$, both $3\%$ less than the same quantities from GONG. The amplitudes from MDI then decreased from 2008 to 2010 by $\sim 8\%$, while the mode widths for the same period of time do not show a clear trend.

The mode amplitudes and widths from HMI nicely merge and continue the solar cycle trend shown by MDI, and are similar to those from GONG. The mode amplitudes obtained from HMI decreased from 2010 to 2013 by $\sim 7\%$ while the mode widths increased by $\sim 1.5\%$. 

In comparison with studies using global mode analysis (\textit{e.g.}, \cite{Salabert06,Garcia13}), the mode parameters from local analysis in this work show a smaller variation over the solar cycle. This is mostly due to our analysis being limited to the central area of the Sun, which is usually outside the activity belt. For a complete picture of solar-cycle variations and a better comparison, we plan to analyze the temporal variations of the mode amplitudes and widths at different latitudes.

The decrease in amplitudes and increase in widths with increase in MAI (bottom panels of Figure 1) are consistent with previous results from global and local analysis. The long-term non-magnetic variations in GONG amplitudes from around mid-2004 to mid-2008, mentioned earlier, introduce large scatter in the values of the mode amplitudes in the range MAI $<$ 8 G. After excluding the data points from GONG for this period, we find the linear correlation coefficient between amplitude and MAI is $-0.80$, and between width and MAI is 0.86. The correlation coefficients for MDI amplitudes and widths are $-0.58$ and 0.49, respectively. This is consistent with lower values of temporal variations with solar activity in the amplitudes and widths from MDI in comparison with those from GONG, discussed above. The correlation coefficients for HMI amplitudes and widths are $-0.70$ and 0.51, respectively. Note however that the HMI results so far only sample a small range in MAI.

According to some predictions and solar activity proxies (\textit{e.g.}, \cite{Lockwood12}, \url{http://solarscience.msfc.nasa.gov/predict.shtml}, \url{http://obs.astro.ucla.edu/150_data.html}), we are almost at the maximum of cycle 24, or may even have reached it. If the predictions are right, than the current solar activity maximum is certainly much lower than that of cycle 23, and the mode parameters clearly reflect this. Due to the lower magnetic activity, the mode amplitudes and widths in this study show consistently smaller variation, about two and three times, respectively, in the rising phase of cycle 24 as compared with the declining phase of cycle 23. It must be noted, however, that solar activity in the rising phase starts at higher latitudes and thus the equatorial part of the Sun, where these comparisons were made, exhibits less magnetic activity during the rising phase than the declining phase of the cycle. The actual differences are likely to be smaller than reported, if we compare the mode parameters near the equator at the same phase of the two cycles.       

\ack
This work utilizes GONG data obtained by the NSO Integrated Synoptic Program (NISP), managed by the National Solar Observatory, which is operated by AURA, Inc. under a cooperative agreement with the National Science Foundation. The data were acquired by instruments operated by the Big Bear Solar Observatory, High Altitude Observatory, Learmonth Solar Observatory, Udaipur Solar Observatory, Instituto de Astrof\'{\i}sica de Canarias, and Cerro Tololo Interamerican Observatory. SOHO is a mission of international cooperation between ESA and NASA. The HMI data are provided by NASA/SDO and the HMI Science Team.

\end{document}